# Guided-wave frequency doubling in surface periodically poled lithium niobate: competing effects

S. Stivala,[1,2] A. Pasquazi,[1] L. Colace,[1] G. Assanto,[1,*] A. C. Busacca,[2] M. Cherchi,[2] S. Riva-Sanseverino,[2,3] A. C. Cino,[3] and A. Parisi[3]

[1]*Department of Electronic Engineering, and Nonlinear Optics and OptoElectronics Laboratory (NooEL), Consorzio Nazionale Italiano di Struttura della Materia (CNISM), Istituto Nazionale di Fisica Nucleare (INFN), University "Roma Tre," Via della Vasca Navale 84, 00146 Rome, Italy*

[2]*Department of Electrical, Electronic and Telecommunication Engineering, University of Palermo, Viale delle Scienze, 90128 Palermo, Italy*

[3]*Centro per la Ricerca Elettronica in Sicilia (CRES), Via Regione Siciliana 49, 90046 Monreale (PA), Italy*

*\*Corresponding author: assanto@ele.uniroma3.it*

We carried out second-harmonic generation in quasi-phase-matched α-phase lithium niobate channel waveguides realized by proton exchange and surface periodic poling. Owing to a limited ferroelectric domain depth, we could observe the interplay between second-harmonic generation and self-phase modulation due to cascading and cubic effects, resulting in a nonlinear resonance shift. Data reduction allowed us to evaluate both the quadratic nonlinearity in the near infrared as well as the depth of the uninverted domains.

## 1. INTRODUCTION

Despite its extensive use and numerous applications since the advent of electric field assisted periodic poling for quasi-phase-matched (QPM) parametric interactions, lithium niobate (LN) remains one of the most investigated dielectrics for harmonic generation (HG) [1–4]. Several techniques have been developed for periodic poling based on LN ferroelectric properties, including electric field poling and electron bombardment [5–7]. In addition, processes for waveguide fabrication in LN have been mastered. The most successful are those based on titanium in-diffusion and proton exchange (PE) [8,9]. Recently, aiming at the realization of shorter and shorter domain lengths for QPM-HG at lower wavelengths [10], as well as counterpropagating wave mixing [11–14], surface periodic poling (SPP) was introduced as a technique to better control mark-to-space ratios in short-period QPM [10,15,16]. Surface periodic poling relies on overpoling a ferroelectric substrate and yields surface-bound periodic poled patterns with shallow domain depths as compared to the thickness of the treated substrate [15]. This approach, used in conjunction with waveguides of comparable depth, is expected to push forward some of the existing frontiers in backward HG by short-period QPM, presently limited to high-order interactions [17–21].

In this paper, we report the first results on picosecond second-harmonic generation (SHG) in surface periodically poled channel waveguides realized by proton exchange in $z$-cut LN. The experimental results exhibit a linear shift of the SHG resonance with fundamental frequency (FF) excitation in the near infrared. The latter shift, readily interpreted in terms of both quadratic cascading [22–30] from the homogeneous portion of the waveguide and Kerr self-focusing, allowed us to infer the quadratic nonlinearity and the domain depth in PE-SPP LN waveguides.

## 2. SURFACE PERIODICALLY POLED WAVEGUIDES IN LITHIUM NIOBATE

Albeit initially demonstrated a few years ago, SPP of LN has only been recently combined with PE for waveguide fabrication in order to enhance the nonlinear response [31]. The 500 $\mu$m thick $z$-cut substrates of congruent LN were spin-coated with photoresist (1.3 $\mu$m thick) and UV exposed to define a $\Lambda = 16.8$ $\mu$m periodic pattern on the −$z$ surface. The samples were then electric field poled using 1.3 kV voltage pulses over a 10 kV bias in order to overcome the LN coercive field (∼22 kV/mm). To this extent, the −$z$ surface was connected to the ground and the +$z$ to the high-voltage (HV) electrode, ensuring a uniform field distribution by means of an electrolyte gel [15]. Applied voltage and current were monitored via the HV generator–amplifier and an oscilloscope in order to achieve the sample "overpoling" after domain merging, i.e., a complete ferroelectric inversion with the exception of relatively shallow unpoled domains at the −$z$ LN surface under the photoresist pattern (see Fig. 1) [15]. Chemical etching in diluted hydrofluoric acid (HF) allowed to reveal the QPM grating and its 50:50 mark-to-space ratio. For waveguide fabrica-tion, the poled LN was initially coated with a $SiO_2$ layer, where channel openings, from 1 to 7 $\mu$m in width, were defined by standard photolithography and wet etching in HF. Proton exchange was then carried out by dipping the LN sample by the sealed-ampoule method [32] in a solution of benzoic acid and lithium benzoate (3%) in order to obtain low-proton-concentration ($\alpha$ phase [33]) channel waveguides supporting TM-polarized modes. We verified the compatibility between SPP and PE by chemically etching the samples and using scanning electron microscopy. Figure 2(b) is the optical micrograph of a typical QPM channel after HF etching. We also employed the planar waveguides formed on the $+z$ face of the samples for distributed (prism and/or grating) coupling at 632.8 nm and evaluating the extraordinary index profile versus $z$ using the standard WKB technique [34]. This yielded $n_e(z) = 2.2027 + 0.0150 \exp[-(z/2.601)^{1.8355}]$, i.e., a waveguide depth of 2.6 $\mu$m, as shown in Fig. 2(a). Using Sellmeier equations [35] and measured TM effective indices from planar waveguides at $\lambda_{FF} = 1.55$ $\mu$m (single mode, $N_0 = 2.1395$) and $\lambda_{SH} = 775$ nm (two modes, $N_0 = 2.1845$ and $N_1 = 2.1801$), respectively, we derived the profiles $n_e(z) = \Delta n_e(z) + n_e(0)$ with $n_e(0) = 2.1381$ and $\Delta n_e(z) = 0.01185 \exp[-(z/2.601)^{1.8355}]$ at $\lambda_{FF}$ and $n_e(0) = 2.1788$ and $\Delta n_e(z) = 0.01395 \exp[-(z/2.601)^{1.8355}]$ at $\lambda_{SH}$, respectively, to be employed in the interpretation of the data from SHG in the SPP-PE channels.

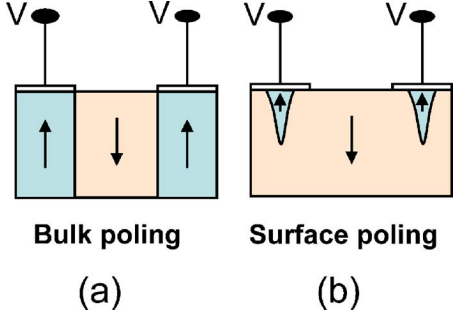

Fig. 1. (Color online) Comparison of (a) bulk poling via electric field and (b) surface poling owing to overpoling.

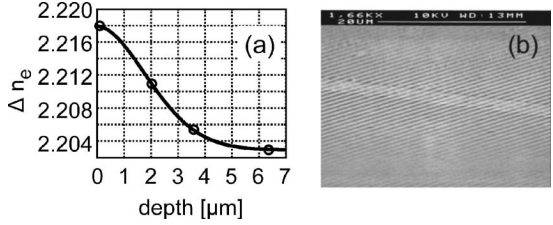

Fig. 2. (a) Extraordinary index profile calculated by inverse WKB from distributed coupling data. The dots are measured effective indices at 632.8 nm; the solid curve is a WKB fit. (b) Scanning electron microphotograph of a typical sample: the PE waveguide is clearly visible on the inverted domain grating.

## 3. SECOND-HARMONIC GENERATION IN SURFACE PERIODIC POLING PROTON-EXCHANGED LITHIUM NIOBATE

Experiments on SHG were carried out to characterize the nonlinear response of the QPM channels. To minimize the detrimental effects of photorefractive damage [36], we employed a source operating at a repetition rate of 10 Hz and producing 25 ps pulses in the near infrared. The source, sketched in Fig. 3, consisted of an optical parametric generator fed by a tunable oscillator and an amplified frequency-doubled Nd:YAG pump at 1.064 $\mu$m. A dispersive element (grating) was introduced in the cavity to effectively narrow the pulse linewidth to <2 $cm^{-1}$ in the interval of wavelength tunability between 1.1 and 1.6 $\mu$m. Following a spectrometer and a single-pulse autocorrelator, the beam was spatially filtered, collimated, and end-fire coupled in the channels through a 20× microscope objective. A $TEM_{00}$ spot of waist $w_0 \sim 3.6$ $\mu$m provided the best in-coupling efficiency. The output fundamental and second-harmonic profiles were imaged with Vidicon and Si-CCD cameras, respectively, using a 63× microscope objective; energy and peak power were measured with Ge and Si photodiodes and a dual-input boxcar averager. All measurements were taken at room temperature with the aid of a Peltier cell and a temperature controller.

Second-harmonic generation measurements were conducted either at a fixed wavelength by varying the FF input power or by scanning the FF wavelength, ratioing the second-harmonic output to the fundamental input to obtain the conversion efficiency. At variance with [37], pulse walk-off and group-velocity dispersion were uninfluent, owing to the picosecond pulse duration and sample length $L = 1$ cm. Typical results versus wavelength are shown in Fig. 4 for an FF input peak power of 2 kW: even though the data (symbols) are plotted versus wavelength rather than detuning, a sinclike shape is apparent and in substantial agreement with the expected one for cw excitation (solid curve). The full width at half maximum is $\sim 1$ nm, yielding an effective QPM grating length of 1 cm, coincident with the sample length. Despite the low efficiency, the latter result confirmed a good uniformity of the PE poled sample in both QPM grating and waveguide parameters [38].

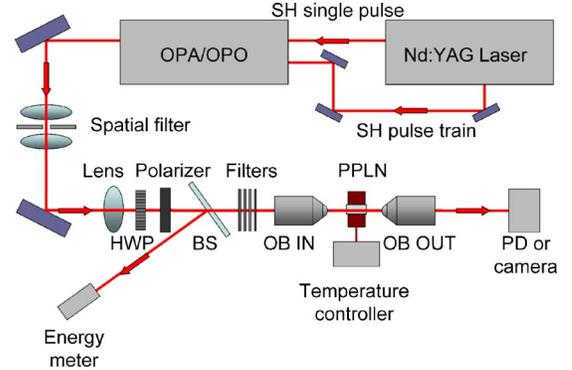

Fig. 3. (Color online) Setup for SHG measurements. HWP, half-wave-plate; BS, beam splitter; OB, microscope objective; PD, photodetector.

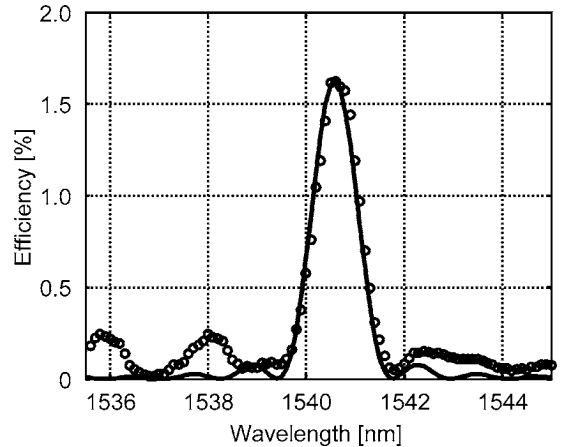

Fig. 4. Measured (open circles) and predicted (solid curve) SHG conversion efficiency versus FF wavelength for a launched peak power of 2 kW.

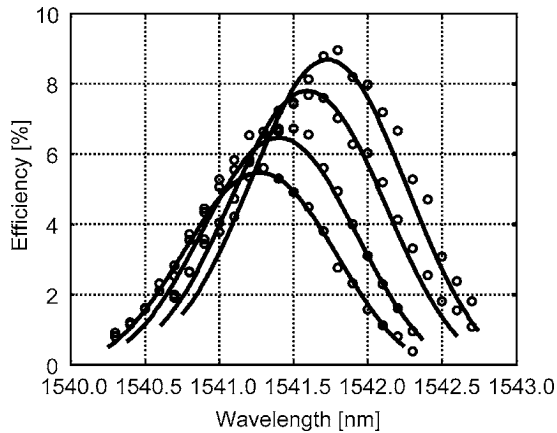

Fig. 5. SHG resonance shift in wavelength for increasing peak excitations 7.4, 8.9, 11.2, and 12.7 kW from left to right, respectively. The experimental values (open circles) are numerically interpolated using Eq. (2) (solid curves).

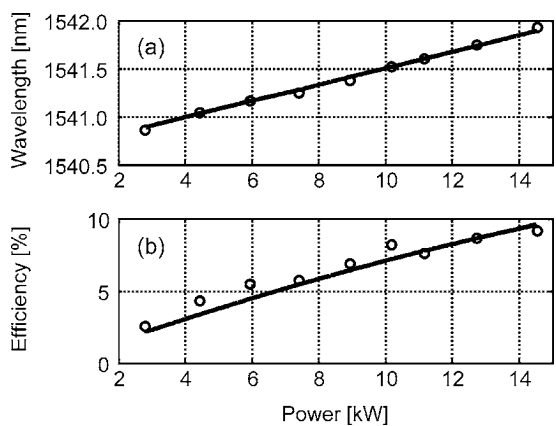

Fig. 6. (a) Peak SHG wavelength shift and (b) maximum conversion efficiency versus FF peak power.

By repeating the SHG scan at higher peak powers we observed that the wavelength of maximum conversion (the FF resonance wavelength for SHG) shifted with excitation, as visible in Fig. 5. Otherwise stated, when performing a power scan at fixed FF wavelength, the conversion efficiency first increased with power (as expected) and then decreased. By adjusting $\lambda_{FF}$, conversely, we observed a linear dependence in maximum conversion to second harmonic (SH) [see Fig. 6(b)], as well as in peak SHG wavelength [see Fig. 6(a)].

## 4. MODEL AND DATA REDUCTION

Considering the measured conversion efficiencies, using the standard model of an ideal (first-order, 50:50 mark-to-space) QPM channel waveguide and taking into account the measured modal overlap integral (input and output), coupling, and Fresnel losses—as well as material absorption at FF and SH—we could estimate an effective nonlinear coefficient $\sim 2$ orders of magnitude smaller than previously reported $d_{33}$ values in LN [35]. Since no appreciable reduction in nonlinearity can be attributed to PE in $\alpha$-phase waveguides [33], a low (effective) nonlinear response can be ascribed to the limited depth of the noninverted domains as compared to the depth of the waveguide; hence, to the limited overlap between (FF and SH) guided modes and the transverse size of the periodic ferroelectric (nonlinear) grating. In fact, for the waveguide depth evaluated in the previous section and the acquired transverse distributions of TM eigenmodes at FF and SH (see Fig. 7), domains extending into $z$ for less than $1-2$ $\mu$m from the surface would greatly limit the pertinent effective area. The latter would imply light propagation and QPM in a partially inverted medium, with detrimental effects on SHG and a nonlinear phase shift due to cascading under phase mismatch [22,23].

Therefore, in order to quantitatively interpret the experimental data, we took into account not only the periodically poled transverse region providing first-order QPM, but also the homogeneously inverted portion of the waveguides extending in $z$ below the actual domains. This contribution is not expected to alter the peak conversion efficiency as the SH generated in the completely inverted region is not phase matched, but it should play an important role in the phase coherently accumulated by the FF in propagation, therefore, on the wavelength for maximum conversion [22,24,29]. A cascaded $\chi^{(2)}:\chi^{(2)}$ process in phase mismatch can mimic a $\chi^{(3)}$ response, giving rise to an equivalent Kerr response (i.e., an index change) that can combine with or even dominate the cubic nonlinearity of the bulk material itself [23] and shift the SHG resonance [37]. Since the QPM grating extended well beyond the channel width (Fig. 2), no particular care needs be adopted along the other transverse coordinate ($y$), where the domains can be assumed infinitely wide.

For monochromatic TM-polarized first-order guided-wave modes at FF and SH, we write the electric field $E$ of an $x$-propagating eigenmode as

$$E(x,y,z;t) = \frac{1}{2}E_u(x,y,z;t)\exp(i\omega_u t - i\beta_u x) + \text{c.c.},$$

$$E_u = \sqrt{\frac{2\eta_0}{N_u \int_{-\infty}^{\infty}\int_{-\infty}^{\infty}|e_u(y,z)|^2 dy dz}}\, u(x)e_u(y,z), \quad (1)$$

with $e_u(y,z)$ the dimensionless transverse profile, $u(x,t)$ the slowly varying amplitude (in space and time) measured in $\sqrt{W}$, $\eta_0$ the vacuum impedance, $N_u$ the effective refractive index, and $\beta_u = \omega_u N_u/c$ the guided-wave propagation constant. In the presence of a rectangular QPM grating with limited-depth inverted domains and an intrinsic third-order response, assuming a negligible contribution from higher-order modes at SH (non-quasi-phase-matched to the FF fundamental mode) and a first-order perturbation, the coupled-mode equations for SHG take the form

$$w_x + \frac{w_t}{c_w} = -w^*v[\chi_1 \exp(-i\Delta\beta_1 x) + i\chi_2 \exp(-i\Delta\beta_2 x)]$$
$$-in_2\frac{2\pi}{\lambda_{\text{FF}}}(f_{ww}|w|^2 + 2f_{wv}|v|^2)w - \frac{\alpha_w}{2},$$

$$v_x + \frac{v_t}{c_v} = w^2[\chi_1 \exp(i\Delta\beta_1 x) - i\chi_2 \exp(i\Delta\beta_2 x)]$$
$$-in_2\frac{4\pi}{\lambda_{\text{FF}}}(2f_{vw}|w|^2 + f_{vv}|v|^2)v - \frac{\alpha_v}{2}, \quad (2)$$

with $w(x,t)$ and $v(x,t)$ being the amplitudes [$u(x,t)$ in Eq. (1)] of fundamental and second-harmonic fields, respectively; $c_w$ and $c_v$ ($\alpha_w$ and $\alpha_v$) being the group velocities (linear absorption coefficients) at FF and SH, respectively; and $n_2$ being the nonresonant intensity-dependent Kerr coefficient [39]. The third-order nonlinear effects, self- and cross-phase modulation, are weighed by the overlap integrals $f_{jk}$:

$$f_{jk} \equiv \frac{\int_{-\infty}^{+\infty}\int_{-\infty}^{+\infty}|e_j(y,z)|^2|e_k(y,z)|^2 dy dz}{\int_{-\infty}^{+\infty}\int_{-\infty}^{+\infty}|e_j(y,z)|^2 dy dz \int_{-\infty}^{+\infty}\int_{-\infty}^{+\infty}|e_k(y,z)|^2 dy dz}$$
$$(j,k=v,w). \quad (3)$$

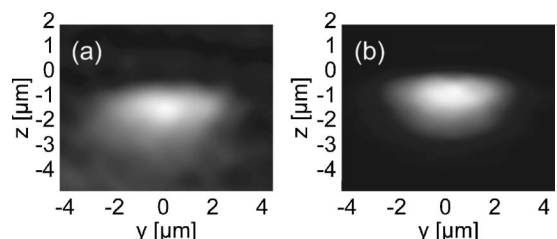

Fig. 7. (a) FF and (b) SH output intensity modal distributions as imaged with a Vidicon tube and a CCD camera, respectively.

In Eq. (2), we also distinguished the SHG contributions from unpoled ($\chi_2$, $\Delta\beta_2 = \beta_v - 2\beta_w$) and poled ($\chi_1$, $\Delta\beta_1 = \Delta\beta_2 - k_G$) regions of the waveguide [40], respectively, governed by mismatches $\Delta\beta_i$ ($i=1,2$) and nonlinear coefficients $\chi_i$ ($i=1,2$), with $k_G = 2\pi/\Lambda$ the wave-vector contribution of the QPM grating with period $\Lambda$. The constants $\chi_1$ and $\chi_2$ can be expressed as

$$\chi_i = d_{\text{eff}_i}\gamma_i\sqrt{\frac{8\pi^2\eta_0 f_{\text{SHG}}}{\lambda_{\text{FF}}^2 N_w^2 N_v}}, \quad (4)$$

with $f_{\text{SHG}}$ the SHG overlap integral

$$f_{\text{SHG}} \equiv \frac{\left[\int_{-\infty}^{+\infty}\int_{-\infty}^{+\infty} e_v^*(y,z)e_w^2(y,z)dy dz\right]^2}{\left[\int_{-\infty}^{+\infty}\int_{-\infty}^{+\infty}|e_w(y,z)|^2 dy dz\right]^2 \int_{-\infty}^{+\infty}\int_{-\infty}^{+\infty}|e_v(y,z)|^2 dy dz}, \quad (5)$$

$d_{\text{eff}_1} = 2d_{33}/\pi$, $d_{\text{eff}_2} = d_{33}$, and $\gamma_i$ parameters accounting for the waveguide section

$$\gamma_1 = \frac{\int_{-\infty}^{+\infty}\int_{Z_0}^{+\infty} e_v^*(y,z)e_w^2(y,z)dy dz}{\int_{-\infty}^{+\infty}\int_{-\infty}^{+\infty} e_v^*(y,z)e_w^2(y,z)dy dz},$$

$$\gamma_2 = \frac{\int_{-\infty}^{+\infty}\int_{-\infty}^{Z_0} e_v^*(y,z)e_w^2(y,z)dy dz}{\int_{-\infty}^{+\infty}\int_{-\infty}^{+\infty} e_v^*(y,z)e_w^2(y,z)dy dz} = 1 - \gamma_1, \quad (6)$$

with $Z_0$ delimiting the domain depth $d$. Note that in Eq. (2) we did not include the cubic nonlinearity due to the QPM grating [41]. In fact, the low conversion efficiency indicates that $d_0$ is quite smaller than the waveguide depth, hence $\chi_2:\chi_1$ and the induced Kerr term $\chi_1:\chi_1$ can be neglected when compared with the cascaded effect stemming from the phase-mismatched SHG in the uniformly poled channel.

While Eqs. (2) model SHG in a partially poled QPM sample, as we deal with picosecond SHG in our low-efficiency samples and focus on the power-dependent shift of the peak conversion wavelength, we can leave out absorption and temporal walk-off as well as the cubic terms depending on the SH field [42,43], reducing the system to

$$w(x,\tau)_x = -w^*v[\chi_1\exp(-i\Delta\beta_1 x) + i\chi_2\exp(-i\Delta\beta_2 x)]$$
$$-i\frac{2\pi}{\lambda_{\text{FF}}}n_2 f_{ww}|w|^2 w,$$

$$v(x,\tau)_x = w^2[\chi_1\exp(i\Delta\beta_1 x) - i\chi_2\exp(i\Delta\beta_2 x)]$$
$$-i\frac{8\pi}{\lambda_{\text{FF}}}n_2 f_{vw}|w|^2 v, \quad (7)$$

having introduced $\tau = t - x/c_w$ and $|w|^2 = w_0^2$ for the $x=0$ input power.

Defining

$$w = W\exp\left(-i\frac{2\pi}{\lambda_{\text{FF}}}n_2 f_{ww}w_o^2 x\right),$$

$$v = V\exp\left(-i\frac{8\pi}{\lambda_{\text{FF}}}n_2 f_{vw}w_o^2 x\right), \quad (8)$$

$$\Delta k_i = \Delta \beta_i + \frac{8\pi}{\lambda_{FF}} n_2 f_{vw} w_o^2 - \frac{4\pi}{\lambda_{FF}} n_2 f_{ww} w_o^2 \quad (i=1,2), \quad (9)$$

the coupled equations for $W(x,\tau)$ and $V(x,\tau)$ take the form

$$W_x = -W^* V[\chi_1 \exp(-i\Delta k_1 x) + i\chi_2 \exp(-i\Delta k_2 x)],$$

$$V_x = W^2[\chi_1 \exp(i\Delta k_1 x) - i\chi_2 \exp(i\Delta k_2 x)]. \quad (10)$$

Finally, since $\Delta k_2 \gg \Delta k_1$ and $\chi_2 \gg \chi_1$, neglecting rapidly oscillating terms such as $\sin[(\Delta \beta_2 - \Delta \beta_1)x]$,

$$W_{xx} + i\Delta k_2 W_x \cong -\chi_2^2(|W|^2 - |V|^2)W. \quad (11)$$

In the small conversion limit and after defining the FF phase $\phi$ as in $W = w_o g(\tau) \exp(-i\phi)$ with $g(\tau)$ the FF temporal profile, Eq. (11) admits the solution [23]

$$\phi = -\frac{\Delta k_2 x}{2}\left(\left[\sqrt{1 + \frac{4\chi_2^2 w_o^2 g(\tau)^2}{(\Delta k_2)^2}} - 1\right]\right) \approx -\frac{\chi_2^2 w_o^2 g(\tau)^2}{\Delta \beta_2}x,$$

for low powers $w_o^2 \ll \Delta k_2 / \chi_2^2$; in the same limit $\Delta \beta_2 \gg (8\pi/\lambda_{FF})n_2 f_{wv} w_o^2$, $(4\pi/\lambda_{FF})n_2 f_{ww} w_o^2$ and the cascading contribution is self-defocusing [22–29].

For the SH output,

$$|v|^2 = \chi_1^2 w_o^4 g^4(\tau) L^2 \operatorname{sinc}^2\left[(\Delta \beta_1 + \Delta C w_o^2 g^2(\tau))\frac{L}{2}\right], \quad (12)$$

with $L$ as the sample length and

$$\Delta C = \frac{8\pi}{\lambda_{FF}} n_2 f_{vw} - \frac{4\pi}{\lambda_{FF}} n_2 f_{ww} + \frac{2\chi_2^2}{\Delta \beta_2}.$$

From Eq. (12), the peak efficiency $\lambda_0$ (the SHG resonant wavelength in the limit of zero FF power) corresponds to $\Delta \beta_1 + \Delta C w_o^2 g^2(\tau) = 0$; therefore, at first order,

$$\lambda = \Delta C \frac{\lambda_o}{k_G} w_o^2 g^2(\tau) + \lambda_o. \quad (13)$$

Expression (13) is consistent with the linear trend we found experimentally (Fig. 6) for a positive $\Delta C$, thereby both cross-phase modulation (with a positive $n_2$) and cascading (with $\Delta \beta_2 > 0$) dominate the $\lambda$ shift. Since no permanent or semipermanent material effects such as hysteresis, memory, or damage could be detected in the samples, other effects (photorefractive, photovoltaic, higher order) potentially contributing to a resonance-wavelength shift were entirely negligible in our waveguides.

We numerically solved Eq. (2) assuming a Gaussian profile for the FF pulses, a propagation length $L = 1$ cm, propagation losses of $\alpha_w = \alpha_v = 0.2$ cm$^{-1}$ at both wavelengths (accounting for both the $\alpha$-phase waveguides and the domain inversion [44,45]), and an input coupling efficiency of 73% consistent with the experimentally measured throughput of 60%. The overlap integrals were calculated from the acquired intensity distributions and shown in Fig. 7, resulting in effective areas $1/f_{ww} = 52.99$ $\mu$m$^2$ and $1/f_{vv} = 23.11$ $\mu$m$^2$ for FF and SH self-phase modulation, respectively; $1/f_{wv} = 44.08$ $\mu$m$^2$ for cross-phase modulation and $1/f_{SHG} = 76.68$ $\mu$m$^2$ for SHG. Since both cubic and quadratic terms contribute to the shift [expression (13)], we adopted Kerr coefficients previously measured by independent methods and reported in literature in order to estimate the pertinent QPM quantities, i.e., nonlinearity $d_{33}$ and the constant $\gamma_1$ related to the depth $d$ of the domains. The latter was extracted from $\gamma_1$ using a modal solver. The index profile along $z$ was reconstructed from the experimental data as described in Section 1; along $y$ we used a generalized Gaussian $\exp(-|y/W_G|^\rho)$ with $\rho$ and $W_G$ parameters to best fit the modal profiles (Fig. 7).

The largest value of $n_2 \approx 10 \times 10^{-20}$ m$^2$/W [46] provided $\gamma_1 = 0.0104$, corresponding to $d_{33} = 16.5$ pm/V and $d \approx 440$ nm. Conversely, the smallest reported $n_2 \approx 5 \times 10^{-20}$ m$^2$/W [47] gave $\gamma_1 = 0.0095$, corresponding to $d_{33} = 18$ pm/V and $d \approx 430$ nm. Therefore, while the domain depth is marginally affected by the size of $n_2$, the quadratic response appears lower than previously reported [38]. Noticeably, neglecting the third-order nonlinearity altogether but keeping the propagation losses, we found $\gamma_1 = 0.0088$, with $d_{33} = 19.5$ pm/V and $d \approx 420$ nm. These results are perfectly consistent with expression (13).

## 5. CONCLUSIONS

We carried out the first experimental demonstration of frequency doubling in proton-exchanged lithium niobate channel waveguides quasi-phase-matched via surface periodic poling. Despite a good mark-to-space ratio of 50:50, we found that a limited depth of the noninverted ferroelectric domains not only limited the overall conversion efficiency to the second harmonic, but also induced a non-negligible quadratic cascading. The latter combined with the inherent cubic response of the crystal to yield an appreciable SHG resonance shift in wavelength. The experimental results confirm the compatibility of $\alpha$-phase proton-exchanged waveguides with surface periodic poling even for periods exceeding 16 $\mu$m but pinpoint the drawback of rather shallow domains as compared with the waveguide index profile. Work is in progress to optimize the PE-SPP technology and achieve deeper ferroelectric domains.

## ACKNOWLEDGMENTS

This work was funded by the Italian Ministry for Scientific Research through PRIN 2005098337 and, in part, through project APQ RS-19 (CIPE 17/2002, BCNanolab).